\documentclass[twoside,twocolumn,english,aps,prl,superscriptaddress,floatfix]{revtex4}
\usepackage[T1]{fontenc}
\usepackage[latin1]{inputenc}
\usepackage{graphicx}

\makeatletter

\newcommand{\amev}[0]{$A$ MeV{}}

\usepackage{babel}
\makeatother
\begin{document}

\preprint{This line only printed with preprint option}

\title{A study of nuclear stopping in central symmetric nuclear collisions
at intermediate energies}

\author{C.~Escano-Rodriguez}
\affiliation{ GANIL, CEA et IN2P3-CNRS, B.P.~55027, F-14076 Caen Cedex, France.}
\author{D.~Durand}
\affiliation{ LPC, IN2P3-CNRS, ENSICAEN et Université, F-14050 Caen Cedex, France.}
\author{A.~Chbihi}
\affiliation{ GANIL, CEA et IN2P3-CNRS, B.P.~55027, F-14076 Caen Cedex, France.}
\author{J.D.~Frankland}
\affiliation{ GANIL, CEA et IN2P3-CNRS, B.P.~55027, F-14076 Caen Cedex, France.}
\author{M.L.~Begemann-Blaich}
\affiliation{ Gesellschaft für Schwerionenforschung mbH, D-64291 Darmstadt, Germany.}
\author{R.~Bittiger}
\affiliation{ Gesellschaft für Schwerionenforschung mbH, D-64291 Darmstadt, Germany.}
\author{B.~Borderie}
\affiliation{ Institut de Physique Nucl\'eaire, IN2P3-CNRS, F-91406 Orsay Cedex,  France.}
\author{R.~Bougault}
\affiliation{ LPC, IN2P3-CNRS, ENSICAEN et Université, F-14050 Caen Cedex, France.}
\author{J.-L.~Charvet}
\affiliation{ DAPNIA/SPhN, CEA/Saclay, F-91191 Gif sur Yvette Cedex, France.}
\author{D.~Cussol}
\affiliation{ LPC, IN2P3-CNRS, ENSICAEN et Université, F-14050 Caen Cedex, France.}
\author{R.~Dayras}
\affiliation{ DAPNIA/SPhN, CEA/Saclay, F-91191 Gif sur Yvette Cedex, France.}
\author{E.~Galichet}
\affiliation{ Institut de Physique Nucl\'eaire, IN2P3-CNRS, F-91406 Orsay Cedex, France.}
\affiliation{ Conservatoire National des Arts et Métiers, F-75141 Paris Cedex 03.}
\author{D.~Guinet}
\affiliation{ Institut de Physique Nucléaire, IN2P3-CNRS et Universit\'e F-69622 Villeurbanne, France.}
\author{P.~Lautesse}
\affiliation{ Institut de Physique Nucléaire, IN2P3-CNRS et Universit\'e F-69622 Villeurbanne, France.}
\author{A.~Le Fèvre}
\affiliation{ Gesellschaft für Schwerionenforschung mbH, D-64291 Darmstadt, Germany.}
\author{R.~Legrain}
\thanks{deceased}
\affiliation{ DAPNIA/SPhN, CEA/Saclay, F-91191 Gif sur Yvette Cedex, France.}
\author{N.~Le~Neindre}
\affiliation{  Institut de Physique Nucl\'eaire, IN2P3-CNRS, F-91406 Orsay Cedex, France.}
\author{O.~Lopez}
\affiliation{ LPC, IN2P3-CNRS, ENSICAEN et Université, F-14050 Caen Cedex, France.}
\author{J.~\L ukasik}
\affiliation{ Gesellschaft für Schwerionenforschung mbH, D-64291 Darmstadt, Germany.}
\author{U.~Lynen}
\affiliation{ Gesellschaft für Schwerionenforschung mbH, D-64291 Darmstadt, Germany.}
\author{L.~Manduci}
\affiliation{ LPC, IN2P3-CNRS, ENSICAEN et Université, F-14050 Caen Cedex, France.}
\author{W.F.J.~Müller}
\affiliation{ Gesellschaft für Schwerionenforschung mbH, D-64291 Darmstadt, Germany.}
\author{L.~Nalpas}
\affiliation{ DAPNIA/SPhN, CEA/Saclay, F-91191 Gif sur Yvette Cedex, France.}
\author{H.~Orth}
\affiliation{ Gesellschaft für Schwerionenforschung mbH, D-64291 Darmstadt, Germany.}
\author{M.~P\^arlog}
\affiliation{ National Institute for Physics and Nuclear Engineering, RO-76900 Bucharest-M\u{a}gurele, Romania.}
\author{M.~F.~Rivet} 
\affiliation{ Institut de Physique Nucl\'eaire, IN2P3-CNRS, F-91406 Orsay Cedex,  France.} 
\author{E.~Rosato}
\affiliation{ Dipartimento di Scienze Fisiche e Sezione INFN, Università di Napoli ``Federico II'', I-80126 Napoli, Italy.}
\author{R.~Roy}
\affiliation{ Laboratoire de Physique Nucléaire, Université Laval, Québec, Canada.}
\author{A.~Saija}
\affiliation{ Gesellschaft für Schwerionenforschung mbH, D-64291 Darmstadt, Germany.}
\author{C.~Schwarz}
\affiliation{ Gesellschaft für Schwerionenforschung mbH, D-64291 Darmstadt, Germany.}
\author{C.~Sfienti}
\affiliation{ Gesellschaft für Schwerionenforschung mbH, D-64291 Darmstadt, Germany.}
\author{B.~Tamain}
\affiliation{ LPC, IN2P3-CNRS, ENSICAEN et Université, F-14050 Caen Cedex, France.}
\author{W.~Trautmann}
\affiliation{ Gesellschaft für Schwerionenforschung mbH, D-64291 Darmstadt, Germany.}
\author{A.~Trczinski}
\affiliation{ Soltan Institute for Nuclear Studies, Pl-00681 Warsaw, Poland.}
\author{K.~Turz\'o}
\affiliation{ Gesellschaft für Schwerionenforschung mbH, D-64291 Darmstadt, Germany.}
\author{E.~Vient}
\affiliation{ LPC, IN2P3-CNRS, ENSICAEN et Université, F-14050 Caen Cedex, France.}
\author{M.~Vigilante}
\affiliation{ Dipartimento di Scienze Fisiche e Sezione INFN, Università di Napoli ``Federico II'', I-80126 Napoli, Italy.}
\author{C.~Volant}
\affiliation{ DAPNIA/SPhN, CEA/Saclay, F-91191 Gif sur Yvette Cedex, France.}
\author{J.P.~Wieleczko}
\affiliation{ GANIL, CEA et IN2P3-CNRS, B.P.~55027, F-14076 Caen Cedex, France.}
\author{B.~Zwieglinski}
\affiliation{ Soltan Institute for Nuclear Studies, Pl-00681 Warsaw, Poland.}
\collaboration{INDRA and ALADIN collaborations}
\noaffiliation

\begin{abstract}
Nuclear stopping has been investigated in central symmetric nuclear
collisions at intermediate energies. Firstly, it is found that the
isotropy ratio, $R_{iso}$, reaches a minimum near the Fermi energy
and saturates or slowly increases depending on the mass of the system
as the beam energy increases. An approximate scaling based on the
size of the system is found above the Fermi energy suggesting the
increasing role of in-medium nucleon-nucleon collisions. Secondly,
the charge density distributions in velocity space, \textbf{$dZ/dv_{\Vert}$}
and \textbf{$dZ/dv_{\bot}$}, reveal a strong memory of the entrance
channel and, as such, a sizeable nuclear transparency in the intermediate
energy range. Lastly, it is shown that the width of the transverse
velocity distribution is proportional to the beam velocity.
\end{abstract}

\date{$Revision: 1.2 $ $Date: 2005/01/21 09:24:34 $}

\maketitle
The study of transport phenomena in nuclear reactions at intermediate
energies is of major importance in the understanding of the fundamental
properties of nuclear matter (for an introduction see\cite{DST:review}).
The comparison of the predictions of the microscopic transport models
(see for instance \cite{Ohnishi,QMD:main-ref,BNV:main-ref,BOB:main-ref1,Ono92:amd-12c+12c-29amev})
with experimental data can help improve our knowledge of the basic
ingredients of such models: namely the nuclear equation of state and
as such, the in-medium properties of the nucleon-nucleon interaction.
In this context, it is mandatory to test the different models over
a large systematics in system size and incident energy. Among the
different issues that can be addressed in such a framework, the question
of the thermalization of the system (in particular the damping of
the momentum distribution) in strongly dissipative collisions is among
the most \textbf{}debated in view of its connection with the search
for a phase transition of the liquid-gas like type\cite{Borderie:PT,D'Agostino:2002my}. 

In this paper, we investigate this problem by studying the stopping
power in nuclear reactions at intermediate energies. We take advantage
of the high quality of the data collected for a large variety of
systems with the INDRA multidetector both at GANIL \textbf{}(see
for instance \textbf{}\cite{I9-Mar97,I17-Pla99,I19-Met00,I46-Bor02,I39-Hud03}\textbf{)}
and at GSI \cite{I42-Luk02,I47-Lef04}. Symmetric systems with total
sizes between 80 and 400 mass units and with incident energies between
25 and 100\amev~ have been considered. Event selection is performed
as follows: in view of the experimental detection thresholds in the
backward part of the detection device, we have only considered charged
products which are emitted in the forward velocity space in the centre-of-mass
of the reaction. In this way the effect of the thresholds are minimized
as well as the associated distortions on the global variables considered
in the following. The selection criterium as far as the quality of
the data is concerned was the following: events were retained if the
total detected charge, $Z_{tot}$ (neutrons were not detected) in
the forward hemisphere of the centre-of-mass was larger than 90\%
of half the total charge of the considered system. Contrarily to the
study of the FOPI data presented in \cite{Fopi04:fopi-stopping},
no attempt has been made to symmetrize the data.

\begin{figure}
\includegraphics[%
  width=0.90\columnwidth]{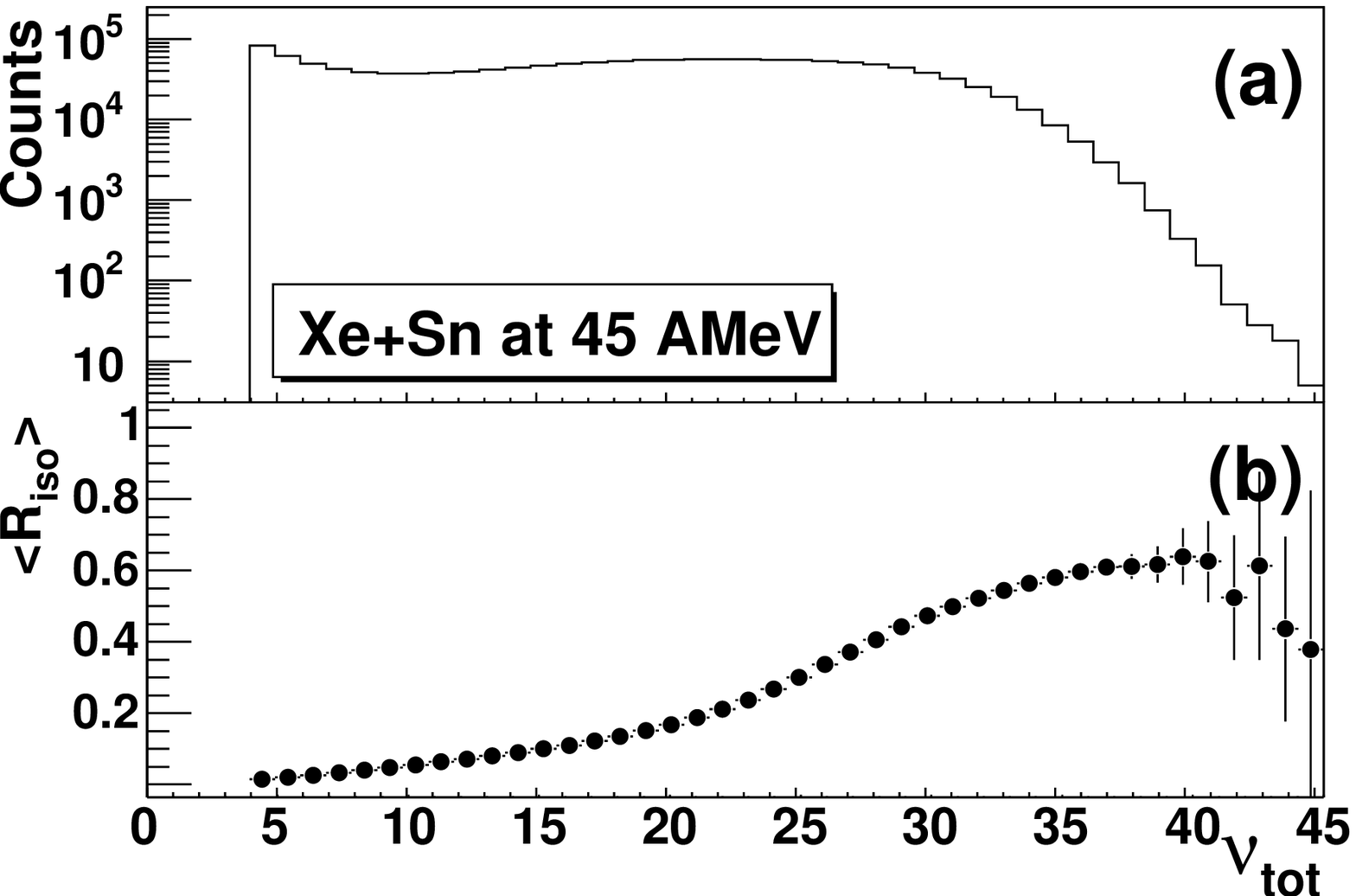}

\includegraphics[%
  width=0.90\columnwidth]{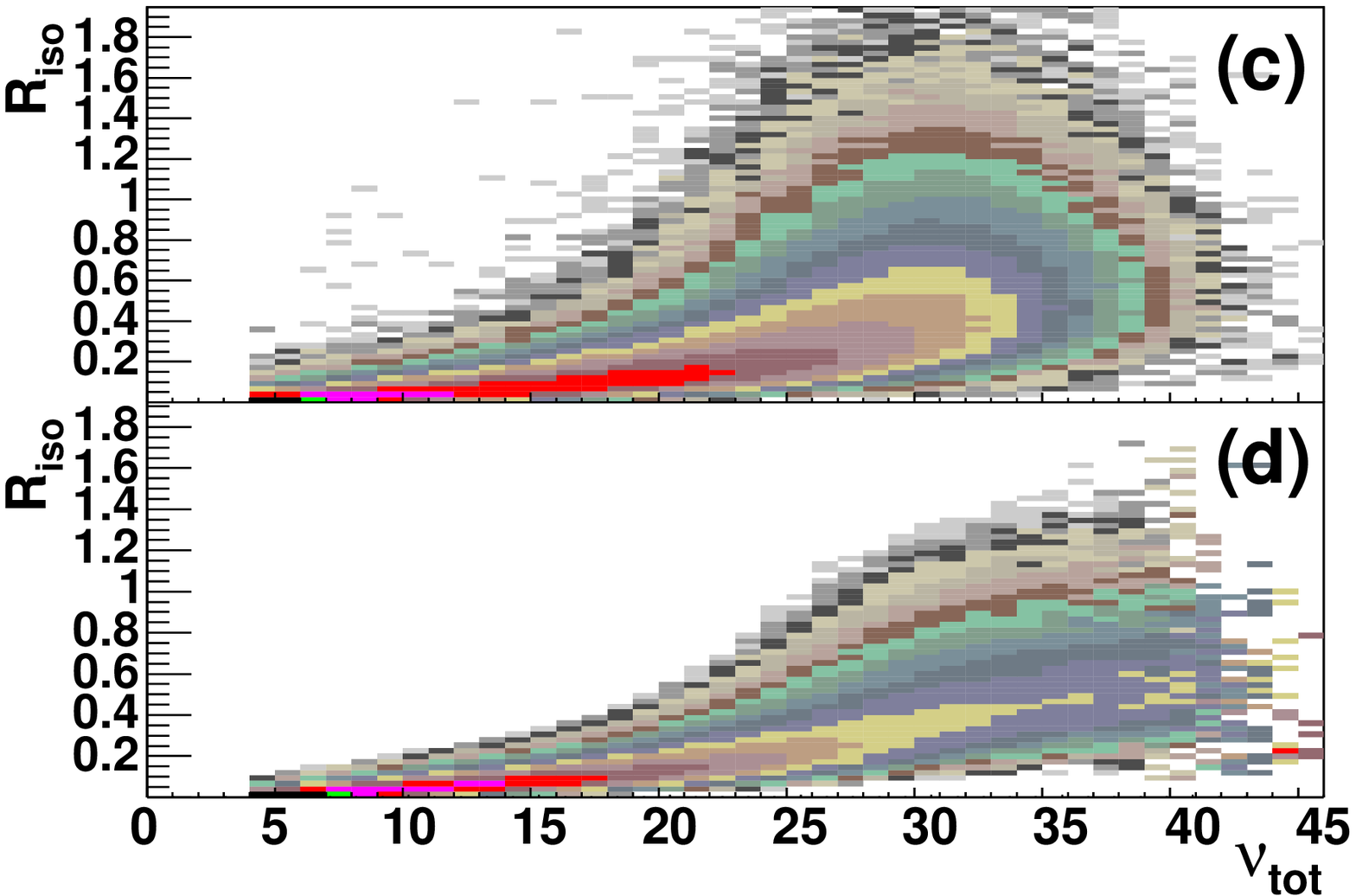}

\caption{Data for Xe+Sn collisions at 45\amev : (a) Distribution of total
charged particle multiplicity $\nu_{tot}$ . (b) Evolution of the
mean isotropy ratio $R_{iso}$~as a function of the total multiplicity,
$\nu_{tot}$. (c) Bi-dimensional plot showing the correlation between
the isotropy ratio and the total multiplicity. Different intensities
(in log scale) represent different numbers of measured events. (d)
As (c), but after renormalising the number of events in each bin according
to the evolution of the cross-section with the multiplicity from (a).
\label{fig1}}
\end{figure}

\begin{figure}[!t]
\includegraphics[%
  width=1.0\columnwidth,
  keepaspectratio]{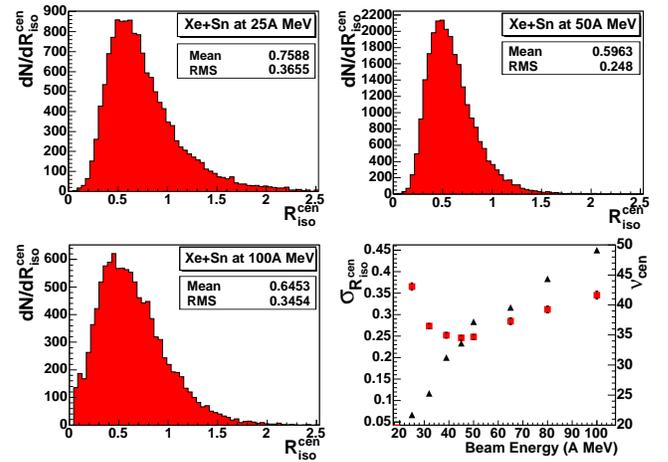}

\caption{Distribution of $R_{iso}^{cen}$ ~in Xe+Sn central collisions. The
incident energy, mean value and variance of $R_{iso}^{cen}$ ~are
indicated in each panel. The last panel shows the evolution of the
RMS (squares and left scale) while triangles and the right scale correspond
to $\nu_{cen}$ defined in the text. Note that both scales do not
start from zero.\label{fig2}}
\end{figure}

In the following, we concentrate on central collisions. This raises
the question of the selection of those events on criteria which do
not induce auto-correlations. Since we are interested in quantities
involving both the transverse and longitudinal directions, it is not
suitable to use vector variables such as, for instance, the transverse
energy \cite{I10-Luk97,I17-Pla99}. Thus, a scalar variable is needed
and the most natural one is the multiplicity of detected charged products,
$\nu_{tot}$. In that sense, a minimum bias selection is used. We
now consider the isotropy ratio, $R_{iso}$ , defined as :

\begin{equation}
R_{iso}=\frac{\sum E_{\bot}}{2\sum E_{//}}\label{eq:1}\end{equation}
where $E_{\bot}$ ( $E_{//}$ ) is the c.m. transverse (parallel)
energy and the sum runs over all products with the above-mentioned
selection. We first consider the evolution of $R_{iso}$ as a function
of $\nu_{tot}$ . Another possibility was to consider the evolution
as a function of the particle multiplicity emitted in the forward
direction but this does not affect the conclusions. The four panels
in Figure \ref{fig1} illustrate the selection method of the central
collisions. Figure \textbf{}\ref{fig1}a shows the distribution of
$\nu_{tot}$ while Figure \ref{fig1}b displays the evolution of the
mean value of $R_{iso}$ as a function of $\nu_{tot}$. The mean value
of \textbf{$R_{iso}$} displays a S-like shape and reaches an asymptotic
value hereafter noted \textbf{$R_{iso}^{cen}$,} and its corresponding
mean value \textbf{$<R_{iso}^{cen}>$}, for the largest values of
\textbf{$\nu_{tot}$.} Strong fluctuations are observed for extreme
values of \textbf{$\nu_{tot}$} where the statistics is very low (typically
less than 10 events, see Figure \textbf{}\ref{fig1}a)\textbf{.} In
the following, we calculate the asymptotic value of $R_{iso}$ by
considering the mean value of $R_{iso}$ for events with a multiplicity
larger than $\nu_{cen}$ (these are plotted as a function of the incident
energy as triangles (right scale) in the last panel of Figure \ref{fig2}).
\textbf{$\nu_{cen}$} is defined as the multiplicity of charged products
at which $R_{iso}$ reaches its asymptotic value. \textbf{}Figures
\ref{fig1}c and \ref{fig1}d \textbf{}show the \textbf{}bi-dimensional
correlation between $\nu_{tot}$ and $R_{iso}$. It is worth \textbf{}noting
from Fig. \ref{fig1}c that the largest fluctuations of $R_{iso}$
are not associated with the largest values of the multiplicity. This
could suggest that the selection of central events by means of $\nu_{tot}$
is inappropriate. Let us discuss briefly this point. A first effect
to consider is the influence of the statistics of the multiplicity
distribution : this has been accounted for in Fig. \ref{fig1}d by
dividing the contents of each bin by the number of events with the
corresponding multiplicity obtained from Fig. \ref{fig1}a. \textbf{}The
large fluctuations observed for intermediate values of the multiplicity
have disappeared demonstrating that this effect is to a large extent
due to statistics.

Another argument in favour of the present selection of central collisions
concerns the velocity-dependent charge densities \cite{I18-Lec00,I43-Gal03}
(defined later in this paper). Such distributions (not shown here
except for central collisions, see Figure \ref{fig4}) do not change
anymore for multiplicities  $\nu_{tot}\geq\nu_{cen}$. This shows that for multiplicities
larger than $\nu_{cen}$, the events have the same kinematical characteristics
and that they only differ by the sorting variable, namely the multiplicity.
We call these events \char`\"{}central events\char`\"{} : they correspond
to roughly \textbf{}1\% of the reaction cross-section and are the
only events considered in the following.

The present study has been undertaken for a variety of systems in
the intermediate energy range. Let us first consider the nearly \textbf{}symmetric
Xe+Sn system. Figure \ref{fig2} shows examples of $R_{iso}^{cen}$
distributions between 25 and 100\amev~incident energy. The distributions
are not gaussian-like and the most probable values do not drastically
change and remain close to 0.5-0.6 as the beam energy increases. The
evolution of the mean value is thus to a large extent governed by
the evolution of the width of the distributions. These latter are
plotted in the last panel (squares and left scale). A minimum is observed
near 40-50\amev~close to the Fermi energy. Also shown are the values
of $\nu_{cen}$ (triangles) in the last panel. They increase from
about 22 at 25\amev~up to 47 at 100\amev.

\begin{figure}
\includegraphics[%
  width=1.0\columnwidth,
  keepaspectratio]{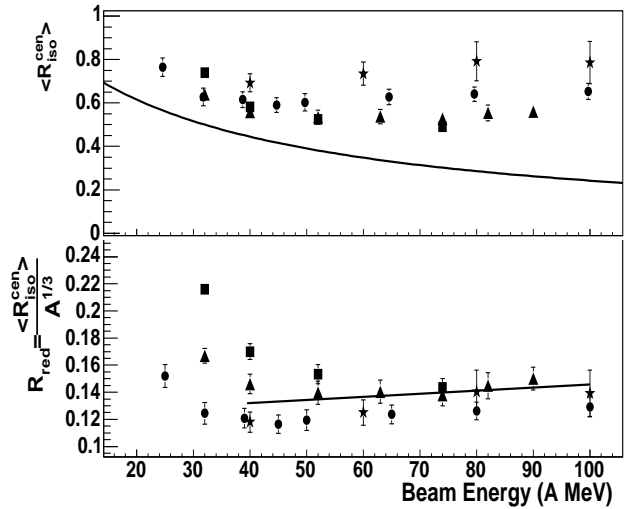}

\caption{Top: Evolution of the mean value of $R_{iso}^{cen}$ (and the associated
uncertainty) as a function of the beam energy for central collisions
and for various systems: (stars) Au+Au, (dots) Xe+Sn, (triangles)
Ni+Ni and (squares) Ar+KCl. The solid line is the value of $R_{iso}$
corresponding to the initial nucleon momentum distributions consisting
of two sharp Fermi spheres separated by a relative velocity corresponding
to the considered beam energy. Bottom: Same as above for the scaled
isotropy ratio: $<R_{iso}^{cen}>/A^{1/3}$. The solid line is a linear
fit to all data points above 40\amev.\label{fig3}}
\end{figure}

Figure \ref{fig3} shows the evolution of $<R_{iso}^{cen}>$ for various
symmetric systems whose total sizes vary from about 80 up to 400 mass
units. The upper panel of the figure shows that, below 40\amev, the
isotropy ratio is nearly independent of the size of the system and
slowly decreases. However, as the beam energy crosses the Fermi energy,
one observes an increase of the isotropy ratio depending on the size
of the system: the larger the size, the larger $<R_{iso}^{cen}>$.
The solid line on Figure \ref{fig3}-top is the value of the isotropy
ratio in the entrance channel due to the Fermi distribution of the
nucleons inside the projectile and the target. Thus, the distance
between the experimental points and the solid line is a ''measure''
of the influence of the dissipation on the isotropy ratio.

The general evolution of the data is interpreted as a transition from
the dominant influence of the mean field at low energy (one-body dissipation
which does not depend on the size of the system) towards the dominance
of in medium nucleon-nucleon collisions at higher energy (two-body
dissipation which depends on the size of the system).

In this case, the key quantity is the ratio between the nucleon mean
free path and the size of the system. This is tentatively put in evidence
in Figure \ref{fig3} (lower panel) where the reduced isotropy ratio
$R_{red}=<R_{iso}^{cen}>/A^{1/3}$ has been plotted for all considered
systems (here $A$ has been arbitrarily taken as half the total mass
of the system as we are dealing with symmetric systems). For medium
and heavy systems (Xe+Sn and Au+Au), the scaling is evidenced for
beam energies larger than 40-50\amev~while for lighter systems (Ar+KCl
and Ni+Ni), it is more and more verified as the beam energy increases.
These features outline the role of the size of the system as far as
the damping of the nucleon momentum distribution is concerned. This
scaling and the rather low values of \textbf{$<R_{iso}^{cen}>$} suggest
that the mean free path of the nucleons inside the medium is rather
long at such incident energies. It is of the same order of magnitude
or even larger than the size of the system. One should also have in
mind surface effects in the sense that even for central collisions,
those nucleons that are more localized in the vicinity of the surface
have little chance to experience hard collisions: All in all, this
means that two-body dissipation is not enough to drive the system
towards thermalization.

Up to now, we have only considered a global variable, $R_{iso}$,
built on an event by event basis. We now consider a variable which
is averaged over all the selected events: namely the velocity-dependent
charge density, $dZ/d\eta_{\Vert}^{red}$ and $dZ/d\eta_{\bot}^{red}$
respectively along the beam axis and along one of the transverse directions.
These quantities are built by considering all charged particles emitted
in the forward centre of mass direction, adding the particles (weighted
with their respective nuclear charges) which are localized in the
same velocity bin. These distributions are well-suited to a comparison
with microscopic transport models, as shown for instance in \cite{I43-Gal03}.
The same distributions have been studied for FOPI data in \cite{Fopi04:fopi-stopping}.
In order to compare easily the same system at different beam energies,
scaled velocities have been used: $\eta=v/v_{cm}$ where $v_{cm}$
is the projectile velocity in the centre-of-mass frame.

\begin{figure*}
\includegraphics[%
  width=0.60\textwidth,
  height=0.55\textwidth,
  keepaspectratio]{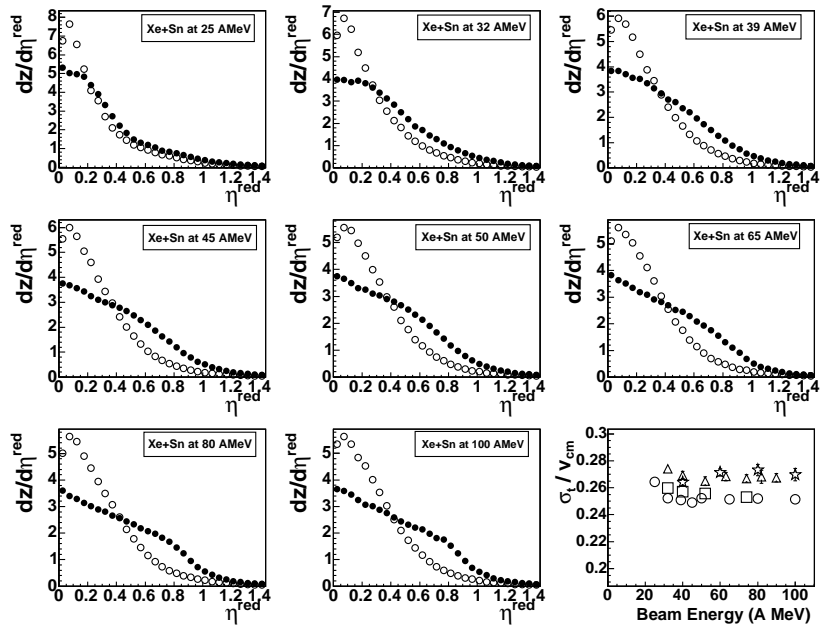}

\caption{Top: $dZ/d\eta_{\Vert}^{red}$ (black points) and $dZ/d\eta_{\bot}^{red}$(open
points) in the parallel and in one of the two transverse directions
for central Xe+Sn collisions for various incident energies indicated
in each panel. The distributions are normalized in such way that the
total charge corresponds approximately to half the total charge of
the sytem. Last panel: Evolution as a function of the incident energy
of the width of the charge density distribution in the transverse
direction, for (stars) Au + Au, (circles) Xe + Sn, (triangles) Ni
+ Ni and (squares) Ar + KCl systems.\label{fig4}}
\end{figure*}

Figure \ref{fig4} shows the evolution of the two densities for the
system Xe+Sn at various incident energies between 25 and 100\amev.
There is a clear correlation between $R_{iso}$ and the behaviour
of the two charge densities. Even at 25\amev, the two distributions
are not identical, \textbf{}which would indicate the occurrence of
full stopping. However, this is the incident energy for which the
two distributions are the closer to each other and consequently, it
corresponds to the largest value of $R_{iso}$. For all the considered
energies, a strong memory of the entrance channel is evidenced: the
widths of $dZ/d\eta_{\Vert}^{red}$ are larger than the widths of
$dZ/d\eta_{\bot}^{red}$. A closer look at the transverse distributions
shows that they are quite similar whatever the incident energy. This
is illustrated in the last panel of Figure \ref{fig4}. The reduced
width of $dZ/d\eta_{\bot}^{red}$ shows a remarkable constancy as
a function of the incident energy indicating that the transverse collective
motion is proportional to the beam velocity. Moreover there is no
significant dependence on the size of the system, the same constancy
is observed for other systems with different sizes : Au + Au, Ni +
Ni and Ar + KCl (see Figure \ref{fig4}\textbf{,} last panel). It
would be interesting to study the origin of this effect in microscopic
transport calculations. 

To conclude, we have studied the behaviour of two observables, the
isotropy ratio $R_{iso}$ and the velocity-dependent charge density
$dZ/d\eta^{red}$in central symmetric nuclear reactions at intermediate
energies. The evolution of $R_{iso}$ as a function of the incident
energy shows a minimum followed by a moderate increase for most studied
systems or a plateau around the Fermi energy for the lightest ones.
Rather low values have been measured indicating that full stopping
is far from being achieved in this energy regime. As far as medium-mass
and heavy systems are considered, a scaling law in terms of the size
of the system is observed suggesting the role of in-medium nucleon-nucleon
collisions in the slow increase of the stopping power. The study of
the velocity-dependent charge density in the parallel and transverse
directions with respect to the beam shows a strong memory of the entrance
channel in the sense that the densities in the parallel and transverse
directions do not coincide suggesting transparency as a key feature
of nuclear reactions at intermediate energies. Last, the width of
the transverse charge density using reduced velocities is invariant
in the entire intermediate energy range.

Our findings concerning the stopping power in central nuclear collisions
are in full agreement with a recent study based on very similar observables
performed by the FOPI Collaboration at higher incident energies up
to the GeV region \cite{Fopi04:fopi-stopping}. Our results as well
as those of FOPI should be compared with the predictions of existing
microscopic transport models. As such, we believe that they constitute
a very severe test of these models over a wide range of system sizes
and incident energies.

\end{document}